\documentclass[conference]{IEEEtran}
\IEEEoverridecommandlockouts
\usepackage{amsmath, amssymb, graphicx, xcolor, multirow, multicol, comment, subcaption, algorithm2e, url, float, etoolbox, adjustbox, pgf, soul, geometry, colortbl, booktabs}
\definecolor{ashgraybase}{HTML}{B2BEB5} 
\definecolor{ashgray25}{HTML}{B2BEB5} 
\colorlet{ashgray20}{ashgray25!75!white} 
\colorlet{ashgray15}{ashgray25!60!white} 
\colorlet{ashgray10}{ashgray25!45!white} 
\colorlet{ashgray5}{ashgray25!30!white}  
\colorlet{ashgray30}{ashgraybase!30!white}

\definecolor{ashgray25}{HTML}{B2BEB5} 
\colorlet{ashgray20}{ashgray25!75!white} 
\colorlet{ashgray15}{ashgray25!60!white} 
\colorlet{ashgray10}{ashgray25!45!white} 
\colorlet{ashgray5}{ashgray25!30!white}  

\title{SOURCE TRACING OF SYNTHETIC SPEECH SYSTEMS THROUGH PARALINGUISTIC PRE-TRAINED REPRESENTATIONS}

\author{
    \IEEEauthorblockN{
        Girish\IEEEauthorrefmark{1}\IEEEauthorrefmark{3}\textsuperscript{\dag},
        Mohd Mujtaba Akhtar\IEEEauthorrefmark{2}\textsuperscript{\dag},
        Orchid Chetia Phukan\IEEEauthorrefmark{3}\textsuperscript{\dag}, 
        Drishti Singh\IEEEauthorrefmark{3}\textsuperscript{\dag}\\
        Swarup Ranjan Behera\IEEEauthorrefmark{4}, 
        Pailla Balakrishna Reddy\IEEEauthorrefmark{5},
        Arun Balaji Buduru\IEEEauthorrefmark{1},
        Rajesh Sharma\IEEEauthorrefmark{6}\IEEEauthorrefmark{7}
    }
    \IEEEauthorblockA{
        \IEEEauthorrefmark{1}\textit{UPES, India},
        \IEEEauthorrefmark{2}\textit{V.B.S.P.U, India},
        \IEEEauthorrefmark{3}\textit{IIIT-Delhi, India}, 
        \IEEEauthorrefmark{4}\textit{Independent Researcher, India},
        \IEEEauthorrefmark{5}\textit{Reliance AI, India}, \\
        \IEEEauthorrefmark{6}\textit{University of Tartu, Estonia},
        \IEEEauthorrefmark{7}\textit{Plaksha university,India}\\
        \texttt{\textcolor{blue}{Correspondence:}mmakhtar.research@gmail.com,orchidp@iiitd.ac.in}
    }
    \thanks{\textsuperscript{\dag} Contributed equally as first authors}
}

\begin{document}

\maketitle

\begin{abstract}
In this work, we focus on source tracing of synthetic speech generation systems (STSGS). Each source embeds distinctive paralinguistic features—such as pitch, tone, rhythm, and intonation—into their synthesized speech, reflecting the underlying design of the generation model. While previous research has explored representations from speech pre-trained models (SPTMs), the use of representations from SPTM pre-trained for paralinguistic speech processing, which excel in paralinguistic tasks like synthetic speech detection, speech emotion recognition has not been investigated for STSGS. We hypothesize that representations from paralinguistic SPTM will be more effective due to its ability to capture source-specific paralinguistic cues attributing to its paralinguistic pre-training. Our comparative study of representations from various SOTA SPTMs, including paralinguistic, monolingual, multilingual, and speaker recognition, validates this hypothesis. Furthermore, we explore fusion of representations and propose \textbf{\texttt{TRIO}}, a novel framework that fuses SPTMs using a gated mechanism for adaptive weighting, followed by canonical correlation loss for inter-representation alignment and self-attention for feature refinement. By fusing TRILLsson (Paralinguistic SPTM) and x-vector (Speaker recognition SPTM), \textbf{\texttt{TRIO}} outperforms individual SPTMs, baseline fusion methods, and sets new SOTA for STSGS in comparison to previous works.
\end{abstract}

\section{INTRODUCTION}
\noindent Advancements in audio manipulation technology have blurred the line between real and synthetic speech. Modern text-to-speech (TTS) and voice conversion (VC) systems can produce highly realistic voices, enabling malicious actors to manipulate speech with remarkable accuracy. This growing challenge highlights the urgent need for reliable methods to detect and attribute synthetic speech. This advancement poses significant risks, as malicious entities can exploit synthetic speech for impersonation, fraud, and misinformation. As such the urgent need for robust synthetic speech detection (SSD) solutions becomes undeniable to safeguard trust in digital communication. As a remedy there has been sufficient research into SDD \cite{ref12, ref14, jung2022aasist}. 
Also, the use of representations from speech pre-trained models (SPTMs) such as Wav2vec2, WavLM, Whisper have captured recent attention within the community as these SPTMs provide performance benefit \cite{martin2022vicomtech,  kawa23b_interspeech, guo2024audio}. 
These SPTMs are either fine-tuned or used as feature extractors for extracting representations. Despite much advancement in synthetic speech detection, most of the previous research has mostly focused on distinguishing real and synthetically generated speech i.e. binary classification, but it is not sufficient to predict and mitigate misuse and improve forensic analysis. \par

To further enhance synthetic speech detection from forensic analysis, it is important to understand the exact tool used to generate the speech and this task is known as Source Tracing of Synthetic Speech Generation Systems (STSGS). It has recently captured attention within the community and plays a crucial role in improving the explainability of detection systems, enforcing accountability,  and developing targeted countermeasures against malicious deepfake applications \cite{SA1,SA2,SA3}. 
Each source (TTS, VC) imprint distinctive paralinguistic features—including pitch, tone, rhythm, and intonation—onto their synthesized speech, mirroring the underlying design principles and processing mechanisms of the respective generation models. \par

As such previous research on STSGS have investigated representations from various state-of-the-art (SOTA) SPTMs \cite{klein24_interspeech, yan2024audiodeepfakeattributioninitial, 10810708, phukan2024investigating} for understanding their capability for capturing such source-specific paralinguistic cues. However, they haven't investigated the usage of representations from SPTM pre-trained for paralinguistic speech processing such as TRILLsson \cite{shor22_interspeech} which have shown SOTA behavior for different paralinguistic tasks including tasks such as synthetic speech detection and speech emotion recognition. In this work, we solve this research gap and explore representations from paralinguistic SPTM for STSGS. \textit{We hypothesize that paralinguistic SPTM representations will be the most effective for STSGS, as their specialized paralinguistic pre-training enables them to capture paralinguistic cues unique to each source more effectively than other SPTM representations}. To test this hypothesis, we conduct a comprehensive comparative study of various SOTA SPTMs, including paralinguistic, monolingual, multilingual, and speaker recognition. Our findings validate our hypothesis. \par

Additionally, inspired by prior research demonstrating performance gains through SPTMs representations fusion in related areas such as synthetic speech detection \cite{chetia-phukan-etal-2024-heterogeneity} and speech emotion recognition \cite{wu2023investigation}, we also explore this direction for STSGS. Phukan et al. \cite{phukan2024investigating} have made the initial exploration for fusion of SPTMs representation for STSGS, however, they have considered only a handful of SPTMs representations, here, in our study, we consider a wide range of SOTA SPTMs representations and also the inclusion of paralinguistic SPTM representations that has been missing and a major drawback in their study. To this end, we introduce \textbf{\texttt{TRIO}} (Ga\texttt{\textbf{T}}ed Canonical Cor\texttt{\textbf{R}}elat\texttt{\textbf{IO}}n Attention Network), a novel framework for fusing SPTMs. \textbf{\texttt{TRIO}} employs a gated mechanism for adaptive weighting of representations, incorporates canonical correlation loss for better alignment between the representations, and utilizes self-attention for enhanced feature refinement. By fusing TRILLsson (a paralinguistic SPTM) with x-vector (a speaker recognition SPTM), \textbf{\texttt{TRIO}} achieves superior performance, outperforming individual SPTMs, baseline fusion techniques, and setting a new SOTA benchmark for STSGS in comparison to previous works.

\noindent \textbf{In summary, the key contributions of this work are as follows:}
\begin{itemize}
    \item We carry out a comprehensive comparative analysis of various SOTA SPTMs representations to understand the capability of paralinguistic SPTM representations for STSGS. We show that representations from TRILLsson achieves the topmost performance amongst all other SPTMs representations.
    \item We introduce a novel framework, \textbf{\texttt{TRIO}} for effective fusion of SPTMs representations. \textbf{\texttt{TRIO}} uses a gated mechanism for adaptive representation weighting, applies canonical correlation loss for improved alignment, and employs self-attention for refined feature enhancement. By fusing TRILLsson and x-vector, \textbf{\texttt{TRIO}} surpasses individual SPTMs and baseline fusion methods, setting a new SOTA benchmark for STSGS compared to prior works.
\end{itemize}
\noindent To make our work more accessible and reproducible, we’ve shared the code and models at \footnote{\url{https://github.com/Helix-IIIT-Delhi/TRIO-Source_Tracing}}.

\section{Speech Pre-trained Representations}
In this section, we present a brief overview of the SPTMs used in our study. Wav2Vec2\footnote{\url{https://huggingface.co/facebook/wav2vec2-base}}  \cite{baevski2020wav2vec}, WavLM\footnote{\url{https://huggingface.co/microsoft/wavlm-base}} \cite{chen2022wavlm}, and Unispeech-SAT\footnote{\url{https://huggingface.co/microsoft/unispeech-sat-base}} \cite{chen2022unispeech} are monolingual SPTMs and we consider their base versions pre-trained on LibriSpeech (960 hours of English). Wav2Vec2 was trained to solve a contrastive learning objective, WavLM was pre-trained for solving masked speech modeling and speech denoising simultaneously while Unispeech-SAT was trained in a multi-task speaker-aware format. Both WavLM and Unispeech-SAT have reported SOTA performance in SUPERB. We consider XLS-R\footnote{\url{https://huggingface.co/facebook/wav2vec2-xls-r-300m}} \cite{babu22_interspeech}, Whisper\footnote{\url{https://huggingface.co/openai/whisper-base}} \cite{radford2023robust}, and MMS\footnote{\url{https://huggingface.co/facebook/mms-1b}} \cite{pratap2024scaling} for multilingual SPTMs. We consider their 300M, 74M, and 1B parameters version for XLS-R, Whisper, and MMS respectively. XLS-R, Whisper, MMS were pre-trained on 128, 96, and over 1400 languages respectively. XLS-R and MMS follows Wav2vec2 architecture and pre-trained in a contrastive learning approach while Whisper is a vanilla transformer encoder-decoder architecture and trained in a multi-task manner. We also consider speaker recognition SPTMs such as x-vector\footnote{\url{https://huggingface.co/speechbrain/spkrec-xvect-voxceleb}} \cite{8461375} and ECAPA\footnote{\url{https://huggingface.co/speechbrain/spkrec-ecapa-voxceleb}} \cite{desplanques2020ecapa} as they have shown its effectiveness for synthetic speech detection \cite{chetia-phukan-etal-2024-heterogeneity} as well as STSGS \cite{phukan2024investigating}. However, Phukan et al. \cite{phukan2024investigating} only considered x-vector in their study and here, in our study, we included, ECAPA, which shows further improvement over x-vector in speaker recognition tasks. Both x-vector and ECAPA are trained on Voxceleb1 + Voxceleb2. As paralinguistic SPTM, we consider TRILLsson \footnote{\url{https://www.kaggle.com/models/google/trillsson}} \cite{shor22_interspeech}. It is a distilled model from the SOTA universal paralinguistic conformer (CAP12). TRILLsson representations shows SOTA performance across various paralinguistic tasks such as speech emotion recognition, synthetic speech detection, speaker recognition, and we use the version with 63M parameters. Additionally, we also add Wav2Vec2-emo\footnote{\url{https://huggingface.co/speechbrain/emotion-recognition-wav2vec2-IEMOCAP}}, a SPTM fine-tuned for SER because SER is inherently a paralinguistic application. Before passing the speech samples to SPTMs, we resample them to 16KHz and extract representations from the last hidden state of the frozen SPTMs by mean pooling. We extract representations of 192 for ECAPA; 512 for x-vector, Whisper (We use its encoder); 768 for Wav2vec2, WavLM, Unispeech-SAT, Wav2vec2-emo; 1024 for TRILLsson; 1280 for XLS-R, MMS.

\section{Modeling}
In this section, we discuss the downstream models used with individual representations followed by the proposed framework, \textbf{\texttt{TRIO}} for fusion of SPTMs representations. We use fully connected network (FCN) and CNN as downstream models as they have preferred by previous research as effective downstream networks \cite{chetia-phukan-etal-2024-heterogeneity, phukan2024investigating}. The CNN model consists of two convolutional blocks that receives SPTMs representations as input with 1D-CNN layers of 128 and 64 filters of kernel size 3 with each 1D-CNN layer followed by maxpooling. Then we flatten the outputs and use a FCN block that consists of two dense layers with 90 and 45 neurons each followed by the final output layer that uses softmax as activation function and outputs probabilities of the source classes. The FCN model follows the same modeling paradigm as used for the FCN block in the CNN model. The number of trainable parameters in FCN models ranges 0.6 to 0.8M while for CNN models, it varies between 0.8 to 1.2M, depending on the input representations dimensionality. 

\begin{figure}[!bt]
    \centering
    \includegraphics[width=0.8\linewidth]{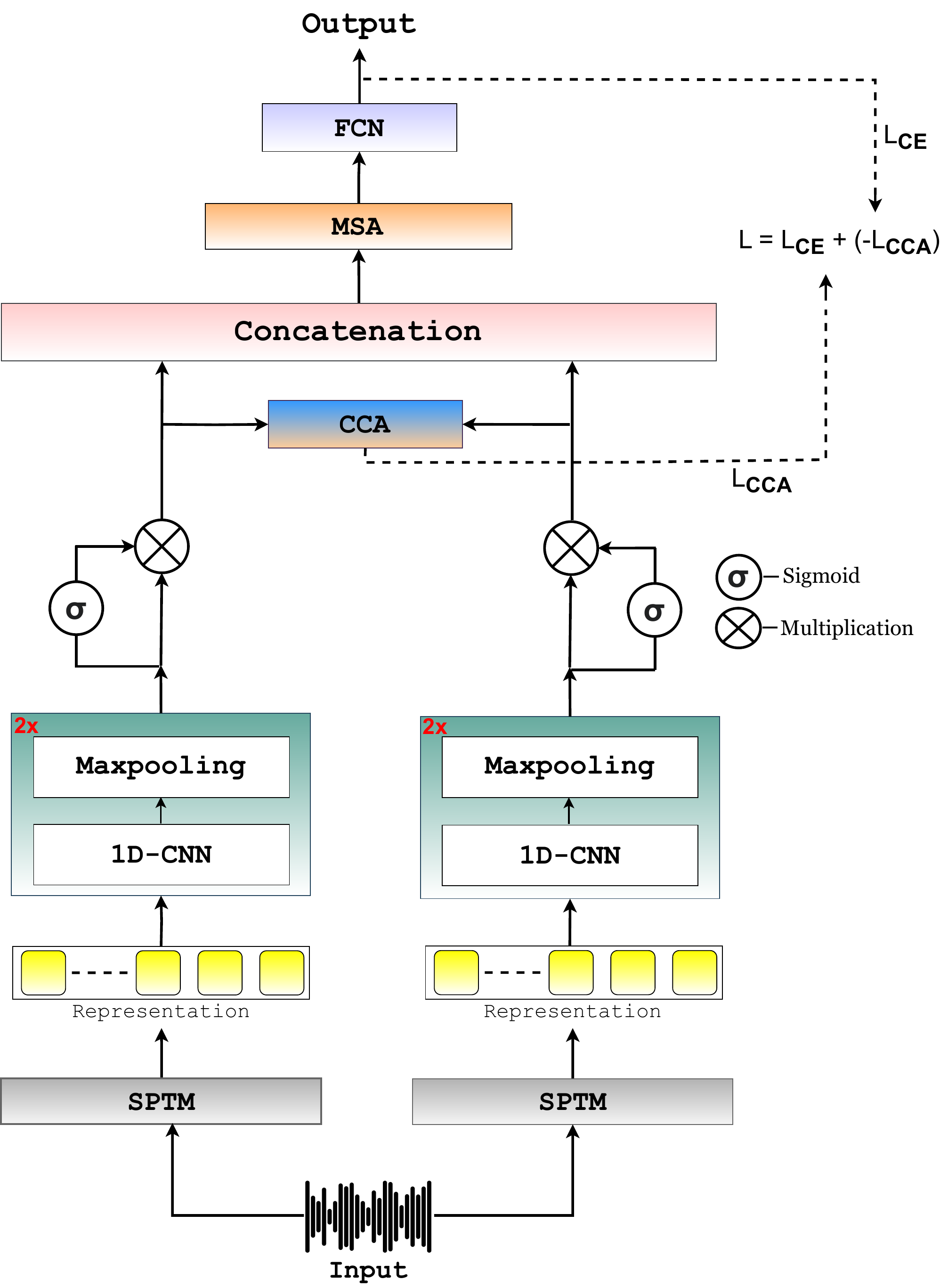}
    \caption{Proposed Framework: \texttt{\textbf{TRIO}}}
    \label{fig:proposed}
\end{figure}

\subsection{\textbf{TRIO}}

The architecture diagram of the proposed framework, \texttt{\textbf{TRIO}} for fusion of SPTMs representations is shown in Figure \ref{fig:proposed}. \texttt{\textbf{TRIO}} leverages a gated mechanism to adaptively weight representations, integrates canonical correlation loss to improve alignment between them, and applies self-attention for more effective feature refinement. First, the SPTMs representations are passed through to two convolutional blocks that uses same modeling as used for individual representational modeling above. Suppose, \( X \) and \( Y \) are features from two SPTMs branches after the flattening them. Then the flattened features are passed through gated mechanism that consists of a sigmoid gate and the outputs are $G_X$ and $G_Y$. After that, we perform element wise-multiplication with the original features, $ \hat{X} = G_X \odot X, \quad \hat{Y} = G_Y \odot Y $ to extract the most relevant features. Next, the refined features are aligned using canonical correlation analysis (CCA) as a novel loss function, which maximizes the correlation between \( \hat{X} \) and \( \hat{Y} \). Higher CCA means better alignment. The CCA loss is formulated as: \[
\mathcal{L}_{CCA} = \text{tr} \left( (\Sigma_{\hat{X}\hat{X}})^{-1/2} \Sigma_{\hat{X}\hat{Y}} (\Sigma_{\hat{Y}\hat{Y}})^{-1/2} \right)
\]

\noindent where $ \Sigma_{\hat{X}\hat{X}} $ and $ \Sigma_{\hat{Y}\hat{Y}} $ are the covariance matrices of $ \hat{X} $ and $ \hat{Y}$, $ \Sigma_{\hat{X}\hat{Y}} $ is the cross-covariance matrix between $ \hat{X} $ and $ \hat{Y} $. $\text{tr}(\cdot) \text{ denotes the trace operation} $. $\mathcal{L}_{CCA}$ ensures that the representations \( \hat{X} \) and \( \hat{Y} \) are maximally correlated, thereby improving their alignment. After aligning the features to a joint representational space, we concatenate the features from the two SPTMs representation networks. Following this, we use a self-attention mechanism, which computes the queries \( Q \), keys \( K \), and values \( V \) as: $ Q = X_{\text{concat}} W_Q, \quad K = X_{\text{concat}} W_K, \quad V = X_{\text{concat}} W_V $ where \( X_{\text{concat}} \) represents the concatenated features from SPTMs representations branches. The attention scores are then computed using the scaled dot-product attention:
\[
\text{Attention}(Q, K, V) = \text{softmax}\left( \frac{Q K^T}{\sqrt{d_k}} \right) V
\]
\noindent Then the features are passed through a FCN block of two dense layers with 90 and 45 neurons followed by a output layer with softmax activation that outputs probabilities. We perform joint optimization with the cross entropy loss $\mathcal{L}_{CCA}$. Finally, the total loss $\mathcal{L}$ is given as: $
\mathcal{L} = \mathcal{L}_{CE} + (- \lambda \cdot \mathcal{L}_{CCA}) $ where \( \lambda \) is a hyperparameter controlling the importance of $\mathcal{L}_{CCA}$. The negative sign before \( \mathcal{L}_{CCA} \) is used because $\mathcal{L}_{CCA}$ is formulated to maximize correlation, while loss functions are typically minimized in optimization. By using a negative sign, we effectively encourage the model to maximize correlation while jointly minimizing $\mathcal{L}_{CE}$. The trainable parameters range from 1.3 to 1.5M.

\section{Experiments}

\subsection{Dataset}
We use two benchmark synthetic speech detection databases: ASVSpoof 2019 (ASV) \cite{wang2020asvspoof} and FAD Chinese Dataset (CFAD) \cite{ma2024cfad}. ASV contains both real and synthetic speech samples from 19 generative systems, recorded at 16 kHz. Real recordings feature diverse speakers with varying accents and speaking styles, while synthetic samples were generated using SOTA VC and TTS methods. We merged the train, validation, and test splits for ASV, resulting in 19 synthetic speech source classes (A01 to A19). We followed 5-fold cross-validation for ASV, with 4 folds used for training and one fold for testing. CFAD is a chinese dataset and features real and synthetic samples from 12 speech synthesis techniques including SOTA TTS and VC systems. We use the official dataset split for training, validating and evaluation of the models. \par

\noindent \textbf{Training and Hyperparameter Details}: The models are trained for 50 epochs with a batch size of 32, utilizing Adam as optimizer and cross-entropy as the loss function. For experiments with \textbf{\texttt{TRIO}}, we kept the value of $\lambda$ fixed as 0.3 throughout the experiments as preliminary exploration yielded optimal results. Dropout and Early stopping are used for mitigating overfitting.

\begin{table}[hbt!]
\centering
\scriptsize
\setlength{\tabcolsep}{3pt}
\renewcommand{\arraystretch}{0.9} 
\begin{tabular}{l|cc|cc|cc|cc}
\toprule
\textbf{PTMs} & \multicolumn{4}{c|}{\textbf{ASV}} & \multicolumn{4}{c}{\textbf{CFAD}} \\
\cmidrule(r){2-5} \cmidrule(l){6-9}
              & \multicolumn{2}{c|}{\textbf{FCN}} & \multicolumn{2}{c|}{\textbf{CNN}} & \multicolumn{2}{c|}{\textbf{FCN}} & \multicolumn{2}{c}{\textbf{CNN}} \\
\cmidrule(r){2-3} \cmidrule(r){4-5} \cmidrule(r){6-7} \cmidrule(l){8-9}
              & \textbf{A} \(\uparrow\) & \textbf{EER} \(\downarrow\) & \textbf{A} \(\uparrow\) & \textbf{EER} \(\downarrow\) & \textbf{A} \(\uparrow\) & \textbf{EER} \(\downarrow\) & \textbf{A} \(\uparrow\) & \textbf{EER} \(\downarrow\) \\
\midrule
\textbf{W2V}    & 
  \cellcolor{ashgray10}46.24 & 
  \cellcolor{ashgray5}15.44  & 
  \cellcolor{ashgray15}63.76 & 
  \cellcolor{ashgray20}6.78  & 
  \cellcolor{ashgray15}51.37 & 
  \cellcolor{ashgray5}24.20  & 
  \cellcolor{ashgray20}76.59 & 
  \cellcolor{ashgray15}9.19   \\ 

\textbf{WV}     & 
  \cellcolor{ashgray10}35.42 & 
  \cellcolor{ashgray10}14.22 & 
  \cellcolor{ashgray10}47.49 & 
  \cellcolor{ashgray10}11.53 & 
  \cellcolor{ashgray5}34.29  & 
  \cellcolor{ashgray5}25.63  & 
  \cellcolor{ashgray10}37.83 & 
  \cellcolor{ashgray5}21.20   \\ 

\textbf{US}     & 
  \cellcolor{ashgray10}45.68 & 
  \cellcolor{ashgray5}23.19  & 
  \cellcolor{ashgray15}56.48 & 
  \cellcolor{ashgray10}10.04 & 
  \cellcolor{ashgray10}45.93 & 
  \cellcolor{ashgray5}34.91  & 
  \cellcolor{ashgray20}73.65 & 
  \cellcolor{ashgray10}17.64   \\ 

\textbf{XR}     & 
  \cellcolor{ashgray15}65.67 & 
  \cellcolor{ashgray10}10.53 & 
  \cellcolor{ashgray20}80.24 & 
  \cellcolor{ashgray20}5.03  & 
  \cellcolor{ashgray15}52.43 & 
  \cellcolor{ashgray10}17.32 & 
  \cellcolor{ashgray20}77.98 & 
  \cellcolor{ashgray15}8.54   \\ 

\textbf{WP}     & 
  \cellcolor{ashgray20}76.92 & 
  \cellcolor{ashgray15}8.88  & 
  \cellcolor{ashgray20}88.43 & 
  \cellcolor{ashgray20}4.61  & 
  \cellcolor{ashgray20}72.17 & 
  \cellcolor{ashgray10}13.85 & 
  \cellcolor{ashgray20}86.41 & 
  \cellcolor{ashgray20}7.91   \\ 

\textbf{MMS}    & 
  \cellcolor{ashgray20}82.21 & 
  \cellcolor{ashgray20}7.90  & 
  \cellcolor{ashgray20}89.29 & 
  \cellcolor{ashgray20}4.17  & 
  \cellcolor{ashgray20}73.54 & 
  \cellcolor{ashgray10}13.51 & 
  \cellcolor{ashgray20}89.78 & 
  \cellcolor{ashgray20}6.95   \\ 

\textbf{XV}     & 
  \cellcolor{ashgray20}89.46 & 
  \cellcolor{ashgray20}5.43  & 
  \cellcolor{ashgray25}96.63 & 
  \cellcolor{ashgray25}2.13  & 
  \cellcolor{ashgray20}76.85 & 
  \cellcolor{ashgray10}11.63 & 
  \cellcolor{ashgray25}92.33 & 
  \cellcolor{ashgray20}4.47   \\ 

\textbf{EP}     & 
  \cellcolor{ashgray20}85.17 & 
  \cellcolor{ashgray20}4.87  & 
  \cellcolor{ashgray25}93.79 & 
  \cellcolor{ashgray20}4.04  & 
  \cellcolor{ashgray20}74.29 & 
  \cellcolor{ashgray10}10.45 & 
  \cellcolor{ashgray20}88.61 & 
  \cellcolor{ashgray20}4.93   \\ 

\textbf{W2V-emo}& 
  \cellcolor{ashgray20}72.59 & 
  \cellcolor{ashgray15}9.65  & 
  \cellcolor{ashgray20}82.29 & 
  \cellcolor{ashgray20}6.32  & 
  \cellcolor{ashgray20}70.53 & 
  \cellcolor{ashgray10}12.96 & 
  \cellcolor{ashgray20}85.39 & 
  \cellcolor{ashgray10}12.51  \\ 

\textbf{T}      & 
  \cellcolor{ashgray25}\textbf{92.53} & 
  \cellcolor{ashgray20}\textbf{4.79}  & 
  \cellcolor{ashgray25}\textbf{97.16} & 
  \cellcolor{ashgray25}\textbf{1.69}  & 
  \cellcolor{ashgray20}\textbf{78.91} & 
  \cellcolor{ashgray15}\textbf{8.63}  & 
  \cellcolor{ashgray25}\textbf{92.81} & 
  \cellcolor{ashgray20}\textbf{3.37}   \\ 

\bottomrule
\end{tabular}
\caption{Accuracy and EER in \%; Abbreviations used: Wav2vec2 (W2V), WavLM (WV), Unispeech (US), XLS-R (XR), Whisper (WP), MMS (MMS), x-vector (XV), ECAPA (EP), Wav2vec2-emo (W2V-emo), TRILLsson (T); The abrreviations used here are kept same for Table \ref{tab:2}}
\label{tab:1}
\end{table}

\begin{table}[hbt!]
\centering
\scriptsize
\setlength{\tabcolsep}{2pt}
\renewcommand{\arraystretch}{1.0} 
\begin{tabular}{l|cc|cc|cc|cc}
\toprule
\textbf{Pairs} & \multicolumn{4}{c|}{\textbf{ASV}} & \multicolumn{4}{c}{\textbf{CFAD}} \\
\cmidrule(r){2-5} \cmidrule(lr){6-9} 
             & \multicolumn{2}{c|}{\textbf{Concat}} & \multicolumn{2}{c|}{\textbf{\texttt{TRIO}}} & \multicolumn{2}{c|}{\textbf{Concat}} & \multicolumn{2}{c}{\textbf{\texttt{TRIO}}} \\
\cmidrule(r){2-3} \cmidrule(r){4-5} \cmidrule(r){6-7} \cmidrule(lr){8-9}
             & \textbf{A}\(\uparrow\) & \textbf{EER}\(\downarrow\) & \textbf{A}\(\uparrow\) & \textbf{EER}\(\downarrow\) & \textbf{A}\(\uparrow\) & \textbf{EER}\(\downarrow\) & \textbf{A}\(\uparrow\)& \textbf{EER}\(\downarrow\) \\
\midrule
W2V + WV      & 
  \cellcolor{ashgray15}94.31 & \cellcolor{ashgray10}7.69 & 
  \cellcolor{ashgray20}95.97 & \cellcolor{ashgray10}7.62 & 
  \cellcolor{ashgray10}88.79 & \cellcolor{ashgray15}4.89 & 
  \cellcolor{ashgray15}94.28 & \cellcolor{ashgray15}4.61 \\
W2V + US      & 
  \cellcolor{ashgray15}92.55 & \cellcolor{ashgray5}8.34 & 
  \cellcolor{ashgray15}94.85 & \cellcolor{ashgray10}7.58 & 
  \cellcolor{ashgray10}85.25 & \cellcolor{ashgray5}9.28 & 
  \cellcolor{ashgray15}91.39 & \cellcolor{ashgray10}8.85 \\
W2V + XR      & 
  \cellcolor{ashgray15}95.64 & \cellcolor{ashgray5}8.73 & 
  \cellcolor{ashgray20}97.14 & \cellcolor{ashgray10}7.69 & 
  \cellcolor{ashgray15}93.86 & \cellcolor{ashgray10}7.95 & 
  \cellcolor{ashgray15}94.13 & \cellcolor{ashgray10}8.63 \\
W2V + WP      & 
  \cellcolor{ashgray15}96.79 & \cellcolor{ashgray10}7.56 & 
  \cellcolor{ashgray20}95.96 & \cellcolor{ashgray10}7.59 & 
  \cellcolor{ashgray15}94.01 & \cellcolor{ashgray5}8.94 & 
  \cellcolor{ashgray15}93.59 & \cellcolor{ashgray10}7.94 \\
W2V + MMS     & 
  \cellcolor{ashgray15}94.60 & \cellcolor{ashgray10}7.62 & 
  \cellcolor{ashgray20}96.28 & \cellcolor{ashgray15}6.28 & 
  \cellcolor{ashgray10}89.97 & \cellcolor{ashgray5}8.55 & 
  \cellcolor{ashgray15}93.54 & \cellcolor{ashgray10}8.51 \\
W2V + XV      & 
  \cellcolor{ashgray15}96.21 & \cellcolor{ashgray10}7.36 & 
  \cellcolor{ashgray20}95.17 & \cellcolor{ashgray10}7.39 & 
  \cellcolor{ashgray10}86.54 & \cellcolor{ashgray10}7.49 & 
  \cellcolor{ashgray10}89.73 & \cellcolor{ashgray10}7.37 \\
W2V + EP      & 
  \cellcolor{ashgray15}93.08 & \cellcolor{ashgray10}6.59 & 
  \cellcolor{ashgray20}95.25 & \cellcolor{ashgray10}7.14 & 
  \cellcolor{ashgray10}86.21 & \cellcolor{ashgray5}9.54 & 
  \cellcolor{ashgray15}92.62 & \cellcolor{ashgray10}7.56 \\
W2V + W2V-emo & 
  \cellcolor{ashgray15}92.85 & \cellcolor{ashgray10}6.48 & 
  \cellcolor{ashgray20}96.64 & \cellcolor{ashgray15}6.59 & 
  \cellcolor{ashgray10}88.67 & \cellcolor{ashgray5}8.58 & 
  \cellcolor{ashgray15}92.47 & \cellcolor{ashgray10}7.58 \\
W2V + T       & 
  \cellcolor{ashgray15}97.50 & \cellcolor{ashgray15}5.86 & 
  \cellcolor{ashgray20}98.21 & \cellcolor{ashgray15}5.08 & 
  \cellcolor{ashgray15}95.85 & \cellcolor{ashgray10}6.08 & 
  \cellcolor{ashgray20}96.23 & \cellcolor{ashgray15}5.89 \\
\midrule
WV + US       & 
  \cellcolor{ashgray10}86.79 & \cellcolor{ashgray10}6.22 & 
  \cellcolor{ashgray10}88.57 & \cellcolor{ashgray15}4.93 & 
  \cellcolor{ashgray5}78.61  & \cellcolor{ashgray5}9.05  & 
  \cellcolor{ashgray10}89.28 & \cellcolor{ashgray10}8.19 \\
WV + XR       & 
  \cellcolor{ashgray10}85.91 & \cellcolor{ashgray15}5.30 & 
  \cellcolor{ashgray10}87.32 & \cellcolor{ashgray15}4.78 & 
  \cellcolor{ashgray15}91.59 & \cellcolor{ashgray5}8.79  & 
  \cellcolor{ashgray15}91.68 & \cellcolor{ashgray10}7.54 \\
WV + WP       & 
  \cellcolor{ashgray15}93.46 & \cellcolor{ashgray10}6.76 & 
  \cellcolor{ashgray20}95.35 & \cellcolor{ashgray15}5.04 & 
  \cellcolor{ashgray15}93.66 & \cellcolor{ashgray5}8.89  & 
  \cellcolor{ashgray20}95.73 & \cellcolor{ashgray10}7.71 \\
WV + MMS      & 
  \cellcolor{ashgray15}90.31 & \cellcolor{ashgray10}6.40 & 
  \cellcolor{ashgray15}92.39 & \cellcolor{ashgray15}4.97 & 
  \cellcolor{ashgray15}90.55 & \cellcolor{ashgray5}8.01  & 
  \cellcolor{ashgray15}93.23 & \cellcolor{ashgray10}8.59 \\
WV + XV       & 
  \cellcolor{ashgray15}94.89 & \cellcolor{ashgray15}5.49 & 
  \cellcolor{ashgray15}94.72 & \cellcolor{ashgray10}4.30 & 
  \cellcolor{ashgray15}90.27 & \cellcolor{ashgray5}8.81  & 
  \cellcolor{ashgray15}93.79 & \cellcolor{ashgray10}8.69 \\
WV + EP       & 
  \cellcolor{ashgray15}93.84 & \cellcolor{ashgray10}5.06 & 
  \cellcolor{ashgray15}94.29 & \cellcolor{ashgray10}5.87 & 
  \cellcolor{ashgray15}93.21 & \cellcolor{ashgray5}9.81  & 
  \cellcolor{ashgray15}93.85 & \cellcolor{ashgray10}7.29 \\
WV + W2V-emo  & 
  \cellcolor{ashgray10}88.69 & \cellcolor{ashgray10}4.99 & 
  \cellcolor{ashgray15}93.51 & \cellcolor{ashgray10}4.29 & 
  \cellcolor{ashgray15}94.67 & \cellcolor{ashgray5}8.63  & 
  \cellcolor{ashgray15}94.89 & \cellcolor{ashgray10}7.63 \\
WV + T        & 
  \cellcolor{ashgray15}95.81 & \cellcolor{ashgray10}4.55 & 
  \cellcolor{ashgray20}95.16 & \cellcolor{ashgray10}4.33 & 
  \cellcolor{ashgray15}95.29 & \cellcolor{ashgray10}7.86 & 
  \cellcolor{ashgray15}95.21 & \cellcolor{ashgray10}7.21 \\
\midrule
US + XR       & 
  \cellcolor{ashgray10}89.28 & \cellcolor{ashgray15}5.36 & 
  \cellcolor{ashgray10}84.61 & \cellcolor{ashgray15}5.23 & 
  \cellcolor{ashgray5}79.20  & \cellcolor{ashgray5}8.11  & 
  \cellcolor{ashgray10}84.62 & \cellcolor{ashgray10}7.06 \\
US + WP       & 
  \cellcolor{ashgray15}91.59 & \cellcolor{ashgray10}6.04 & 
  \cellcolor{ashgray15}93.82 & \cellcolor{ashgray15}5.27 & 
  \cellcolor{ashgray5}81.82  & \cellcolor{ashgray5}9.24  & 
  \cellcolor{ashgray10}85.06 & \cellcolor{ashgray10}7.29 \\
US + MMS      & 
  \cellcolor{ashgray15}90.55 & \cellcolor{ashgray15}5.22 & 
  \cellcolor{ashgray15}92.38 & \cellcolor{ashgray15}4.51 & 
  \cellcolor{ashgray10}89.63 & \cellcolor{ashgray10}7.99 & 
  \cellcolor{ashgray15}91.50 & \cellcolor{ashgray15}5.72 \\
US + XV       & 
  \cellcolor{ashgray15}92.29 & \cellcolor{ashgray15}5.54 & 
  \cellcolor{ashgray20}97.63 & \cellcolor{ashgray15}4.69 & 
  \cellcolor{ashgray10}88.26 & \cellcolor{ashgray5}8.14  & 
  \cellcolor{ashgray15}92.72 & \cellcolor{ashgray15}5.85 \\
US + EP       & 
  \cellcolor{ashgray15}91.97 & \cellcolor{ashgray15}5.66 & 
  \cellcolor{ashgray15}94.27 & \cellcolor{ashgray15}4.76 & 
  \cellcolor{ashgray10}87.72 & \cellcolor{ashgray5}8.39  & 
  \cellcolor{ashgray15}93.50 & \cellcolor{ashgray15}6.53 \\
US + W2V-emo  & 
  \cellcolor{ashgray15}92.32 & \cellcolor{ashgray15}5.49 & 
  \cellcolor{ashgray15}94.93 & \cellcolor{ashgray15}4.47 & 
  \cellcolor{ashgray15}90.28 & \cellcolor{ashgray5}8.06 & 
  \cellcolor{ashgray15}92.85 & \cellcolor{ashgray15}6.25 \\
US + T        & 
  \cellcolor{ashgray15}93.52 & \cellcolor{ashgray15}4.92 & 
  \cellcolor{ashgray20}95.25 & \cellcolor{ashgray15}4.22 & 
  \cellcolor{ashgray15}91.63 & \cellcolor{ashgray10}7.02 & 
  \cellcolor{ashgray15}94.23 & \cellcolor{ashgray15}4.86 \\
\midrule
XR + WP       & 
  \cellcolor{ashgray15}94.81 & \cellcolor{ashgray15}5.06 & 
  \cellcolor{ashgray20}95.53 & \cellcolor{ashgray15}4.11 & 
  \cellcolor{ashgray15}90.62 & \cellcolor{ashgray15}5.49 & 
  \cellcolor{ashgray20}95.36 & \cellcolor{ashgray15}5.31 \\
XR + MMS      & 
  \cellcolor{ashgray15}94.59 & \cellcolor{ashgray15}5.72 & 
  \cellcolor{ashgray20}95.83 & \cellcolor{ashgray15}5.14 & 
  \cellcolor{ashgray15}92.36 & \cellcolor{ashgray10}6.52 & 
  \cellcolor{ashgray15}94.82 & \cellcolor{ashgray15}5.17 \\
XR + XV       & 
  \cellcolor{ashgray15}94.27 & \cellcolor{ashgray15}4.95 & 
  \cellcolor{ashgray20}95.37 & \cellcolor{ashgray15}5.35 & 
  \cellcolor{ashgray15}90.89 & \cellcolor{ashgray15}5.30 & 
  \cellcolor{ashgray15}93.81 & \cellcolor{ashgray15}5.52 \\
XR + EP       & 
  \cellcolor{ashgray15}93.51 & \cellcolor{ashgray15}4.92 & 
  \cellcolor{ashgray15}94.43 & \cellcolor{ashgray15}4.26 & 
  \cellcolor{ashgray15}91.76 & \cellcolor{ashgray15}5.87 & 
  \cellcolor{ashgray15}92.29 & \cellcolor{ashgray15}4.49 \\
XR + W2V-emo  & 
  \cellcolor{ashgray15}93.84 & \cellcolor{ashgray15}5.29 & 
  \cellcolor{ashgray15}94.11 & \cellcolor{ashgray15}4.64 & 
  \cellcolor{ashgray15}93.83 & \cellcolor{ashgray15}5.19 & 
  \cellcolor{ashgray15}94.08 & \cellcolor{ashgray15}4.21 \\
XR + T        & 
  \cellcolor{ashgray15}94.62 & \cellcolor{ashgray15}4.38 & 
  \cellcolor{ashgray20}96.89 & \cellcolor{ashgray15}4.05 & 
  \cellcolor{ashgray15}94.05 & \cellcolor{ashgray15}4.84 & 
  \cellcolor{ashgray20}95.13 & \cellcolor{ashgray15}3.91 \\
\midrule
WP + MMS      & 
  \cellcolor{ashgray15}93.59 & \cellcolor{ashgray15}4.93 & 
  \cellcolor{ashgray15}94.44 & \cellcolor{ashgray15}4.48 & 
  \cellcolor{ashgray15}92.52 & \cellcolor{ashgray10}6.29 & 
  \cellcolor{ashgray15}92.89 & \cellcolor{ashgray15}4.96 \\
WP + XV       & 
  \cellcolor{ashgray15}95.13 & \cellcolor{ashgray15}5.31 & 
  \cellcolor{ashgray20}96.01 & \cellcolor{ashgray15}4.29 & 
  \cellcolor{ashgray15}95.11 & \cellcolor{ashgray15}5.19 & 
  \cellcolor{ashgray20}97.16 & \cellcolor{ashgray15}4.14 \\
WP + EP       & 
  \cellcolor{ashgray15}93.81 & \cellcolor{ashgray15}4.89 & 
  \cellcolor{ashgray15}94.06 & \cellcolor{ashgray15}4.21 & 
  \cellcolor{ashgray15}93.66 & \cellcolor{ashgray15}4.81 & 
  \cellcolor{ashgray15}92.28 & \cellcolor{ashgray15}4.23 \\
WP + W2V-emo  & 
  \cellcolor{ashgray15}94.26 & \cellcolor{ashgray15}4.73 & 
  \cellcolor{ashgray20}95.24 & \cellcolor{ashgray15}4.02 & 
  \cellcolor{ashgray15}92.98 & \cellcolor{ashgray15}4.51 & 
  \cellcolor{ashgray15}94.08 & \cellcolor{ashgray15}3.85 \\
WP + T        & 
  \cellcolor{ashgray15}94.89 & \cellcolor{ashgray15}3.95 & 
  \cellcolor{ashgray20}95.31 & \cellcolor{ashgray20}3.24 & 
  \cellcolor{ashgray15}95.21 & \cellcolor{ashgray15}4.09 & 
  \cellcolor{ashgray20}97.52 & \cellcolor{ashgray20}3.09 \\
\midrule
MMS + XV      & 
  \cellcolor{ashgray15}96.89 & \cellcolor{ashgray15}3.84 & 
  \cellcolor{ashgray20}97.51 & \cellcolor{ashgray20}3.53 & 
  \cellcolor{ashgray15}93.53 & \cellcolor{ashgray15}5.27 & 
  \cellcolor{ashgray15}94.48 & \cellcolor{ashgray15}4.61 \\
MMS + EP      & 
  \cellcolor{ashgray15}95.67 & \cellcolor{ashgray15}3.18 & 
  \cellcolor{ashgray20}96.11 & \cellcolor{ashgray20}2.96 & 
  \cellcolor{ashgray15}91.13 & \cellcolor{ashgray15}3.59 & 
  \cellcolor{ashgray15}92.34 & \cellcolor{ashgray20}3.38 \\
MMS + W2V-emo  & 
  \cellcolor{ashgray15}96.91 & \cellcolor{ashgray15}3.08 & 
  \cellcolor{ashgray20}97.86 & \cellcolor{ashgray20}2.93 & 
  \cellcolor{ashgray15}92.89 & \cellcolor{ashgray15}4.19 & 
  \cellcolor{ashgray15}93.82 & \cellcolor{ashgray20}3.91 \\
MMS + T       & 
  \cellcolor{ashgray15}97.17 & \cellcolor{ashgray15}2.84 & 
  \cellcolor{ashgray20}98.21 & \cellcolor{ashgray20}2.72 & 
  \cellcolor{ashgray15}93.86 & \cellcolor{ashgray15}3.53 & 
  \cellcolor{ashgray15}94.28 & \cellcolor{ashgray20}3.01 \\
\midrule
XV + EP       & 
  \cellcolor{ashgray15}97.16 & \cellcolor{ashgray15}4.24 & 
  \cellcolor{ashgray20}98.04 & \cellcolor{ashgray15}4.15 & 
  \cellcolor{ashgray15}94.22 & \cellcolor{ashgray15}4.21 & 
  \cellcolor{ashgray15}95.68 & \cellcolor{ashgray15}4.03 \\
XV + W2V-emo  & 
  \cellcolor{ashgray15}97.25 & \cellcolor{ashgray15}4.31 & 
  \cellcolor{ashgray20}98.14 & \cellcolor{ashgray15}4.01 & 
  \cellcolor{ashgray15}95.14 & \cellcolor{ashgray15}4.63 & 
  \cellcolor{ashgray20}96.55 & \cellcolor{ashgray15}4.14 \\
\textbf{XV + T} & 
  \cellcolor{ashgray25}\textbf{98.38} & \cellcolor{ashgray25}\textbf{0.36} & 
  \cellcolor{ashgray25}\textbf{99.56} & \cellcolor{ashgray25}\textbf{0.19} & 
  \cellcolor{ashgray25}\textbf{97.28} & \cellcolor{ashgray25}\textbf{1.29} & 
  \cellcolor{ashgray25}\textbf{99.04} & \cellcolor{ashgray25}\textbf{0.95} \\
\midrule
EP + W2V-emo  & 
  \cellcolor{ashgray10}87.61 & \cellcolor{ashgray5}8.53 & 
  \cellcolor{ashgray10}89.93 & \cellcolor{ashgray10}7.21 & 
  \cellcolor{ashgray5}76.28  & \cellcolor{ashgray5}10.64 & 
  \cellcolor{ashgray10}79.94 & \cellcolor{ashgray10}8.28 \\
EP + T        & 
  \cellcolor{ashgray15}92.28 & \cellcolor{ashgray15}2.10 & 
  \cellcolor{ashgray15}94.81 & \cellcolor{ashgray20}1.83 & 
  \cellcolor{ashgray5}79.24  & \cellcolor{ashgray10}7.61 & 
  \cellcolor{ashgray10}82.38 & \cellcolor{ashgray15}5.53 \\
\midrule
W2V-emo + T   & 
  \cellcolor{ashgray15}97.39 & \cellcolor{ashgray15}0.45 & 
  \cellcolor{ashgray20}97.56 & \cellcolor{ashgray20}0.39 & 
  \cellcolor{ashgray15}96.38 & \cellcolor{ashgray15}1.49 & 
  \cellcolor{ashgray20}97.16 & \cellcolor{ashgray20}0.99 \\
\bottomrule
\end{tabular}
\caption{Accuracy and EER in \%}
\label{tab:2}
\end{table}

\subsection{Experimental Results}

\begin{figure}[!h]
    \centering
    \begin{minipage}{0.24\textwidth}
        \centering
        \includegraphics[width=\textwidth]{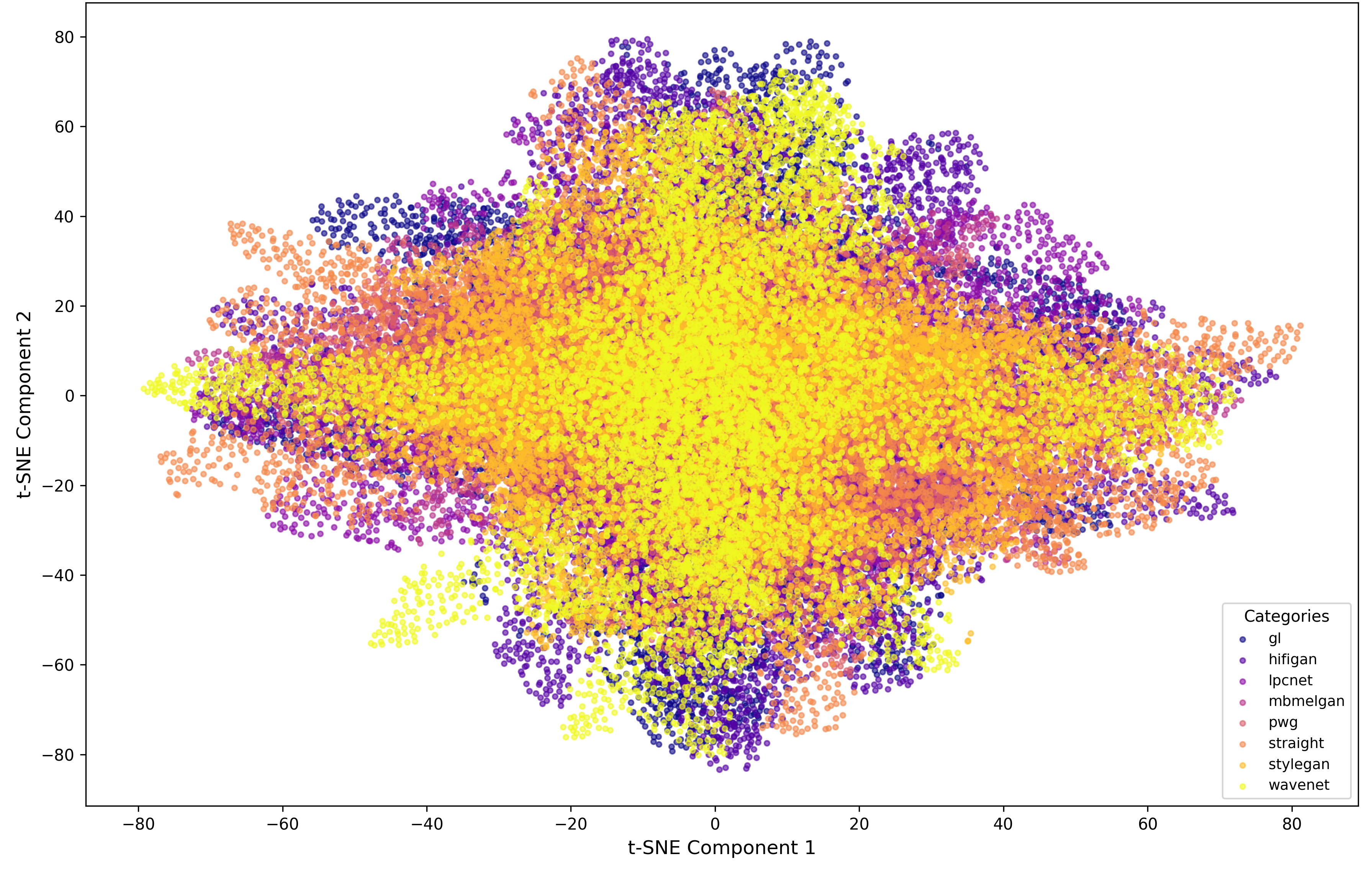} 
        \caption*{(a) MMS}
    \end{minipage}\hfill
    \begin{minipage}{0.24\textwidth}
        \centering
        \includegraphics[width=\textwidth]{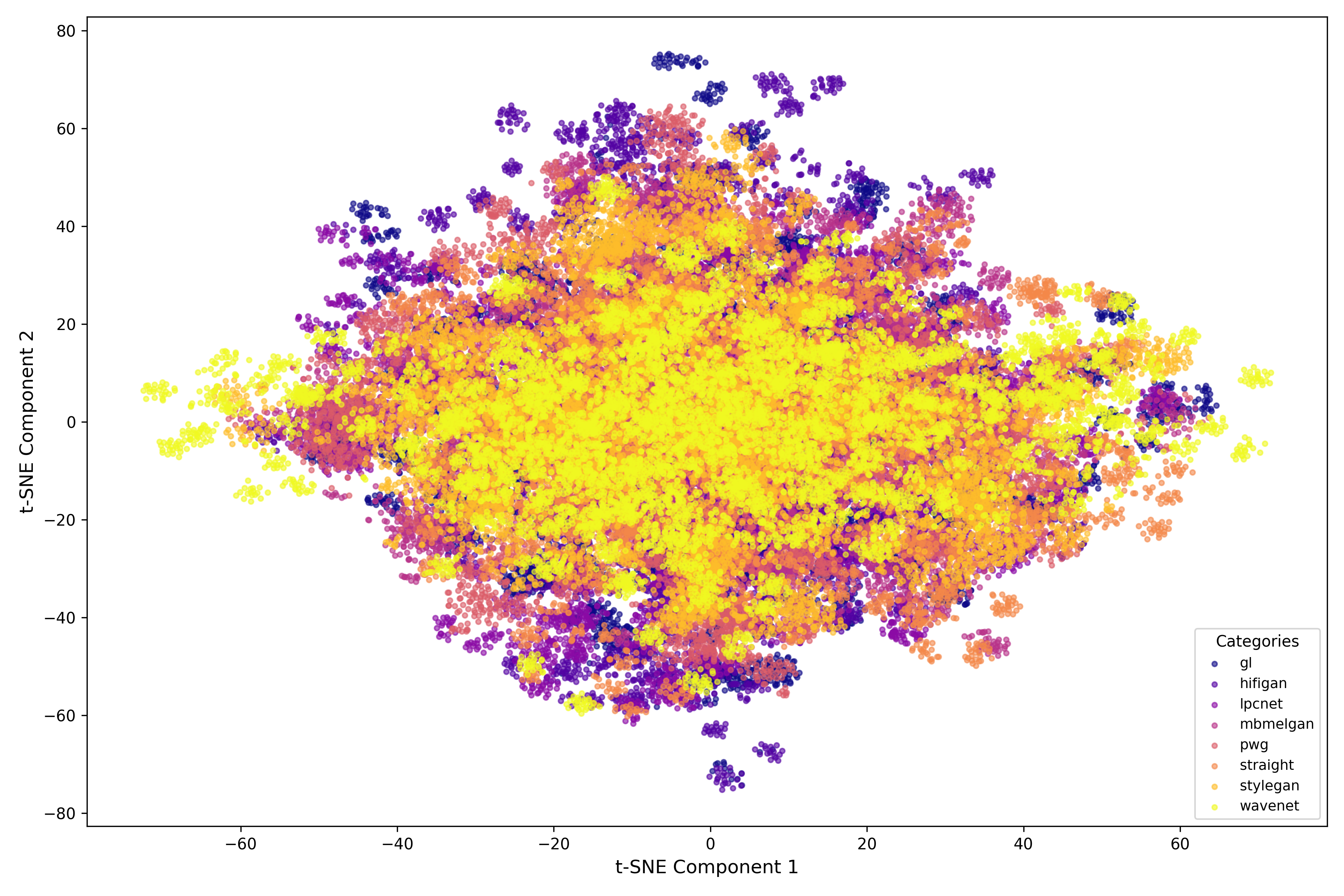} 
        \caption*{(b) TRILLsson}
    \end{minipage}\\[10pt] 

    \caption{t-SNE Plots for CFAD} 
    \label{fig:tsne}
\end{figure}


\begin{figure}[!h]
    \centering
    \includegraphics[width=0.3\textwidth]{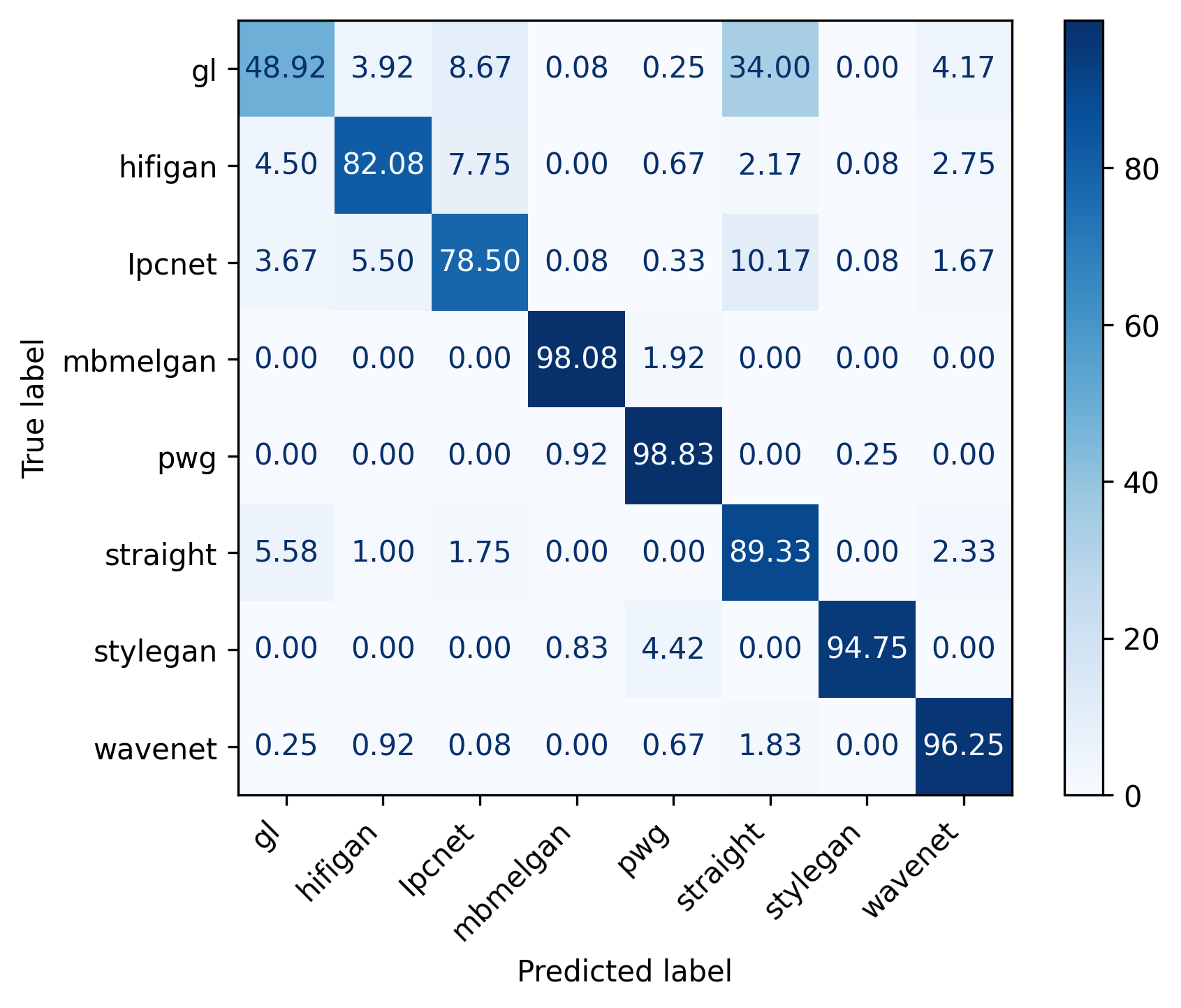}
    \caption{Confusion Matrix for CFAD using \textbf{\texttt{TRIO (x-vector + TRILLsson)}}}
    \label{fig:confusion_matrices}
\end{figure}

Table \ref{tab:1} presents the evaluation scores of downstream networks trained on different SPTMs representations. We use accuracy and equal error rate (EER) as the evaluation metrics following previous research on STSGS \cite{phukan2024investigating}. We report EER by computing the average scores using a one-vs-all approach. For ASV, we report the average of five folds scores and for CFAD, we report the scores obtained in the official evaluation set. Our findings indicate that representations from TRILLsson (paralinguistic SPTM) consistently achieve the highest attribution accuracy and the lowest EER, significantly outperforming other representations. This reinforces their ability to capture source-specific paralinguistic cues, which are crucial for distinguishing synthetic speech sources. This validates our hypothesis that paralinguistic SPTM representations will be the most effective for STSGS attributing to their paralinguistic pre-training. Among all the other SPTMs, speaker recognition SPTMs (x-vector and ECAPA) showed comparatively good performance. This suggests that their pre-training for speaker recognition tasks enhances their ability to capture source-specific cues, contributing to improved performance in STSGS. Additionally, we observe that monolingual SPTMs reports the lowest performance for both the datasets showing its inability to capture source specific cues. Overall, the CNN models showed better performance than its FCN counterparts. We also plot the t-SNE plots of raw representations of MMS and TRILLsson in Figure \ref{fig:tsne}. We observe better cluster across the source classes for TRILLsson and this supports our obtained results and further amplify the credibility of the proposed hypothesis. \par
Table \ref{tab:2} presents the results of fusion of various SPTMs representations. We use concatention-based fusion as the baseline fusion technique. We keep the same network for concatenation-based fusion technique as the proposed framework, \textbf{\texttt{TRIO}}. However, we remove the gated mechansim, CCA loss and self-attention refinement block. We keep the training details as same as for experiments with \textbf{\texttt{TRIO}}. Our results indicate that fusion of representations through \textbf{\texttt{TRIO}} outperforms the baseline fusion technique, demonstrating its effectiveness in integrating diverse SPTM representations. Notably, the best performance across both the datasets was achieved by fusing x-vector and TRILLsson representations using \textbf{\texttt{TRIO}}, highlighting the complementary nature of these representations. Further, we observe that fusion of TRILLsson with speaker recognition and multilingual SPTMs shows comparatively good performance than fusion of monolingual SPTMs with each other. Overall, fusion of SPTMs representations improved performance than the performance with individual representations. We also plot the confusion matrices of CNN trained with \textbf{\texttt{TRIO}} with fusion of x-vector and TRILLsson in Figure \ref{fig:confusion_matrices}.  

\noindent \textbf{Comparison with SOTA}: We compare our best performing model \textbf{\texttt{TRIO}} with x-vector and TRILLsson with previous SOTA work \cite{phukan2024investigating}. They reported accuracy and EER of 98.91\% and 0.26\% on ASV, while for CFAD, they reported 99.01\% and 1.07\%. While we report accuracy and EER: 99.56\% and 0.19\% on ASV, 99.04\% and 0.95\% on CFAD. This top performance shows that our work sets the new SOTA for STSGS. 

\section{Conclusion}
In our study, we show the effectiveness of utilizing paralinguistic SPTMs representations for STSGS. By capturing source-specific paralinguistic cues, these representations outperform representations from various other SOTA SPTMs. Further, we propose \texttt{\textbf{TRIO}}, a novel framework for fusion of representations. By integrating TRILLsson and x-vector representations through \texttt{\textbf{TRIO}}, we show topmost performance surpassing individual SPTMs representations and baseline fusion methods as well as report SOTA results in STSGS compared to previous SOTA work. Our findings serve as a valuable reference for future studies in selecting appropriate SPTMs representations for STSGS and highlight the potential of combining SPTMs representations for further enhancing STSGS.

\bibliographystyle{IEEEtran}
\bibliography{main}

\end{document}